%% file: main.tex
\documentclass[12pt]{scaleai-paper}
\usepackage[square,sort&compress,numbers]{natbib}

\usepackage{mathpazo}

\usepackage[scaled=0.92]{helvet} 
\usepackage{fontawesome5}
\usepackage{subcaption}
\usepackage{hyperref}
\usepackage{tcolorbox}
\usepackage{graphicx}
\usepackage{wrapfig}

\newcommand{\eval}{\textsc{BioRiskEval}\xspace}
\newcommand{\evalgen}{\eval-\textsc{Gen}\xspace}
\newcommand{\evalmut}{\eval-\textsc{Mut}\xspace}
\newcommand{\evalmutprobe}{\eval-\textsc{Mut}-\textsc{Probe}\xspace}
\newcommand{\evalvir}{\eval-\textsc{Vir}\xspace}

\newcommand{\ld}{LD$_{\text{50}}$\xspace}

\usepackage{hyperref}       
\hypersetup{
    colorlinks=true,
	linkcolor=blue,
	filecolor=magenta,      
	urlcolor=blue,
	citecolor=black,
	pdfinfo={
        Title={Best Practices for Biorisk Evaluations on Open-Weight Bio-Foundation Models},
        Subject={ml safety, benchmarks and datasets, adversarial robustness},
        Keywords={ai safety, ml safety, language models, malicious use, misuse, machine unlearning, unlearning, datasets and benchmarks, adversarial robustness, adversarial examples},
    }
}
\usepackage{longtable}
\usepackage{tabularx}
\usepackage[utf8]{inputenc} 
\usepackage[T1]{fontenc}    
\usepackage{hyperref}       
\usepackage{url}            
\usepackage{booktabs}       
\usepackage{amsfonts}       
\usepackage{nicefrac}       
\usepackage{microtype}      
\usepackage{booktabs}
\usepackage{multirow}
\usepackage{amsmath}
\usepackage{xspace}
\usepackage{graphicx}
\usepackage[table]{xcolor}
\usepackage{array}

\usepackage{enumitem}
\setlist[itemize]{leftmargin=*}
\setlist[enumerate]{leftmargin=*}

\usepackage{graphicx}
\usepackage{tcolorbox}

\usepackage{amsmath}
\usepackage{soul}
\usepackage{cleveref}
\usepackage{listings}
\usepackage{tikz}
\usepackage{eso-pic}

\let\svthefootnote\thefootnote
\newcommand\freefootnote[1]{%
  \let\thefootnote\relax%
  \footnotetext{#1}%
  \let\thefootnote\svthefootnote%
}

\input{utils}

\makeatletter
\renewcommand\AB@affilsepx{, \protect\Affilfont}
\makeatother

\title{Best Practices for Biorisk Evaluations on\\Open-Weight Bio-Foundation Models}

\author[1,2$*\dagger$]{Boyi Wei}
\author[1,3$*\dagger$]{Zora Che}
\author[1$\dagger$]{Nathaniel Li}
\author[1]{Udari Madhushani Sehwag}
\author[4]{Jasper Götting}
\author[4]{Samira Nedungadi}
\author[1$\dagger$]{Julian Michael}
\author[1$\dagger$]{Summer Yue}
\author[5]{Dan Hendrycks}
\author[2]{Peter Henderson}
\author[1$\dagger$]{Zifan Wang}
\author[4]{Seth Donoughe}
\author[5]{Mantas Mazeika}

\affil[1]{Scale AI}
\affil[2]{Princeton University}
\affil[3]{University of Maryland}
\affil[4]{SecureBio}
\affil[5]{Center for AI Safety}

\newcommand{\authoremail}{%
  \vspace{-1.5em}
  $^*$\ \textit{Equal Contributions}, 
  $^\dagger$\ \textit{Work done while at Scale AI} \\ \newline
    \faEnvelope\  \texttt{wby@princeton.edu} \quad 
    \faDatabase\  \href{https://github.com/scaleapi/BioRiskEval}{\texttt{github.com/scaleapi/BioRiskEval}} \quad 
    \faGlobe\  \href{https://scale.com/blog/bioriskeval}{\texttt{scale.com/research/bioriskeval}}
}

\begin{document}

\newcommand*\circled[1]{\tikz[baseline=(char.base)]{
            \node[shape=circle,draw,inner sep=1pt] (char) {#1};}}
\newcommand{\watermarktext}{\textbf{Warning: Potentially Harmful Content}}
\newcommand\watermark{%
  \begin{tikzpicture}[remember picture,overlay,scale=3]
    \node[
    rotate=60,
    scale=3,
    opacity=0.3,
    color=red,
    inner sep=0pt
    ]
    at (current page.center) []
    {\watermarktext};
\end{tikzpicture}}%

\maketitle
\authoremail
\input{sections/abstract}

\input{sections/intro}
\input{sections/related_work}

\input{sections/benchmark_intro}

\input{sections/exp_results}

\input{sections/conclusion}

\input{sections/ethics_statement}

\bibliography{main}
\bibliographystyle{abbrvnat}
\newpage
\appendix

\input{appendix/dataset_detail}

\input{appendix/exp_details}


\input{appendix/evo2_model_analysis}

\end{document}

%% file: utils.tex
\definecolor{darkslategray}{rgb}{0.18, 0.31, 0.31}
\definecolor{mybackground}{RGB}{245, 245, 244} 
\definecolor{mytext}{RGB}{0, 0, 0}             
\definecolor{mykeyword}{RGB}{0, 0, 128}        
\definecolor{mycomment}{RGB}{64, 128, 128}     
\definecolor{mystring}{RGB}{128, 0, 0}         
\definecolor{myidentifier}{RGB}{0, 0, 0}       
\definecolor{mynumber}{RGB}{128, 0, 128}       
\definecolor{amethyst}{rgb}{0.6, 0.4, 0.8}

\definecolor{lemon}{RGB}{255,247,0}
\definecolor{maize}{RGB}{250,237,94}
\definecolor{mustard}{RGB}{255,219,89}
\definecolor{ocre}{RGB}{241,103,35}
\definecolor{Tangerine}{RGB}{253,128,8}
\definecolor{framegreen}{RGB}{153, 188, 133}
\definecolor{bggreen}{RGB}{235, 250, 228}
\definecolor{c0}{cmyk}{1,0.3968,0,0.2588} 
\definecolor{c1}{cmyk}{0,0.6175,0.8848,0.1490} 
\definecolor{c2}{cmyk}{0.1127,0.6690,0,0.4431} 
\definecolor{c3}{cmyk}{0.3081,0,0.7209,0.3255} 
\definecolor{c4}{RGB}{164, 16, 52}
\definecolor{orange}{HTML}{E66100}
\definecolor{bluex}{HTML}{0C7BDC}
\definecolor{yellow}{HTML}{FFC20A}
\definecolor{lightpurple}{HTML}{E6E6FA}
\definecolor{lightbluee}{HTML}{e8f4f8}
\definecolor{blush}{rgb}{0.87, 0.36, 0.51}
\definecolor{c5}{HTML}{EE4E4E}
\definecolor{gggggg}{HTML}{EFEFEF}
\definecolor{lightgray}{rgb}{0.83, 0.83, 0.83}
\definecolor{Gred}{RGB}{219, 50, 54}
\definecolor{Ggreen}{RGB}{60, 186, 84}
\definecolor{Gblue}{RGB}{72, 133, 237}
\definecolor{Gyellow}{RGB}{247, 178, 16}
\definecolor{ToCgreen}{RGB}{0, 128, 0}
\definecolor{myGold}{RGB}{231,141,20}
\definecolor{myBlue}{rgb}{0.19,0.41,.65}
\definecolor{myPurple}{RGB}{175,0,124}

\providecommand{\Comments}{1}
\ifnum\Comments>0
\usepackage[colorinlistoftodos,prependcaption,textsize=scriptsize]{todonotes}
\else
\usepackage[disable]{todonotes}
\fi
\usepackage{pifont}
\newcommand{\cmark}{\textcolor{green!60!black}{\ding{51}}}

%% file: sections/abstract.tex
\begin{abstract}
Open-weight bio-foundation models present a dual-use dilemma. While holding great promise for accelerating scientific research and drug development, they could also enable bad actors to develop more deadly bioweapons.
To mitigate the risk posed by these models, current approaches focus on filtering biohazardous data during pre-training. However, the effectiveness of such an approach remains unclear, particularly against determined actors who might fine-tune these models for malicious use.
To address this gap, we propose \eval, a framework to evaluate the robustness of procedures that are intended to reduce the dual-use capabilities of bio-foundation models. \eval assesses models' virus understanding through three lenses, including sequence modeling, mutational effects prediction, and virulence prediction.
Our results show that current filtering practices may not be particularly effective: Excluded knowledge can be rapidly recovered in some cases via fine-tuning, and exhibits broader generalizability in sequence modeling.
Furthermore, dual-use signals may already reside in the pretrained representations, and can be elicited via simple linear probing. 
These findings highlight the challenges of data filtering as a standalone procedure, underscoring the need for further research into robust safety and security strategies for open-weight bio-foundation models.
\end{abstract}


%% file: sections/intro.tex
\section{Introduction}
\label{sec:intro}



The growing capabilities of bio-foundation models have
raised concerns about their potential misuse~\citep{puzis2020increased,urbina2022dual, wang2025call, feldman2025resilient}. This concern is particularly prominent for open-weight models, which allow greater freedom for adversarial modifications, especially when fine-tuned for malicious purposes~\citep{qifine, qi2024evaluating}.

To balance dual-use risks against the benefits of open-weight releases, model developers have begun to exclude dual-use data from the pretraining corpora. For instance, the OpenGenome dataset\footnote{\url{https://huggingface.co/datasets/arcinstitute/opengenome2}}, used to train the Evo model family, intentionally excludes eukaryotic viral sequences~\citep {nguyen2024sequence, brixi2025genome}. As a result, Evo models exhibit poor initial performance in predicting mutational effects on such sequences~\citep{brixi2025genome}. 

Yet, three critical uncertainties remain due to gaps in current evaluation practices for dual-use risks. First, while some initial efforts have been made in examining biorisks in natural language models~\citep{gotting2025virology, yin2025genome, li2024wmdp}, there is no comparable evaluation framework for bio-foundation models, making systematic dual-use risk assessment challenging.
Second, the effectiveness of data filtering for bio-foundation models has not been fully assessed under a threat model where adversaries can fine-tune the model. Data filtering has sometimes been a robust approach for reducing dual-use risks in natural language models, even after some adversarial fine-tuning~\citep{obrien2025deepignorancefilteringpretraining, chen2025enhancing}.
But to date, there has not been a similar assessment in the bio-foundation model context to determine whether the filtered capabilities are trivial to re-learn.
Third, elicitation practices for testing the safety of open-weight bio-foundation models are underexplored. Simple methods including probing have not yet been tried on these models, leaving open the possibility that latent representations still encode the necessary knowledge to enable misuse~\citep{li2024wmdp, jang2025prompt}. 

In this paper, we bridge this gap by evaluating the effectiveness of data filtering as a risk mitigation strategy for bio-foundation models. Our main contributions are as follows:
\begin{itemize}
    \item First, we introduce \eval, an evaluation framework to assess bio-foundation models' capabilities in harmful domains from three perspectives: \textit{Sequence Modeling}, \textit{Mutational Effect Prediction}, and \textit{Virulence Prediction}.
    \item Second, we investigate the effectiveness of data filtering against malicious fine-tuning. To test how cheap and efficient malicious fine-tuning can be done, our analysis focuses on two aspects: the extent to which the fine-tuned model generalizes to unseen sequences, indicating the sample efficiency of the attack; and the minimal compute cost required to recover excluded knowledge.
    \item Third, we analyze whether harmful knowledge is already embedded in latent representations even with the harmful data filtered out during pre-training. By employing the linear probe, we demonstrate that even without fine-tuning, the hidden layer representations from Evo 2 model can be used to predict mutational effects and virulence, achieving comparable performance as the model trained without data filtering.
\end{itemize}

Our work underscores the limitations of current data-exclusion practices as safety and security mechanisms. Future research might improve the robustness of data-exclusion practices, but it is important that policymakers and model developers remain aware of these robustness challenges.

\begin{figure}[t]
    \centering
    \includegraphics[width=\linewidth]{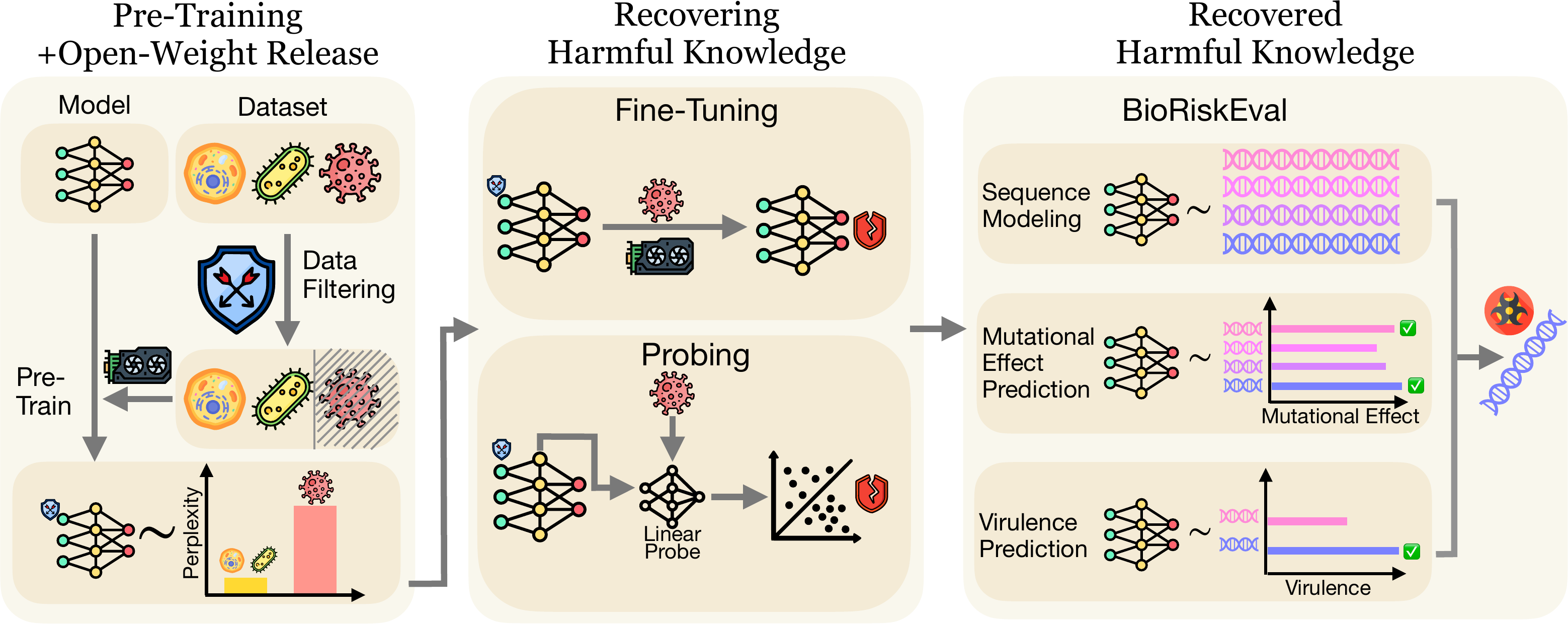}
    \caption{We introduce \eval, a framework for assessing dual-use risk in open-weight bio-foundation models from three perspectives. Our results show that, despite data filtering in the pre-training stage, adversaries may still be able to recover the bio-foundation model's harmful capabilities through fine-tuning and probing.}
    \label{fig:mainfigure}
\end{figure}

%% file: sections/related_work.tex
\section{Related Work}

\input{tables/bfm_tasks}
\paragraph{Bio-foundation models.} Bio-foundation models are large models trained on diverse biological datasets to learn generalized representations that can be adapted to various downstream tasks~\citep{fu2025foundation}. They can be categorized by the type of biomolecular data they process, including \textbf{Genomic Models} (models that operate on genomic sequences, e.g., Enformer~\citep{avsec2021effective}, DNABert~\citep{zhou2023dnabert, ji2021dnabert}, Evo family~\citep{nguyen2024sequence, brixi2025genome}), \textbf{Transcriptomics Models} (models that analyze high-dimensional gene expression data, e.g., Geneformer~\citep{cui2024geneformer}, CellFM~\citep{zeng2025cellfm}), and \textbf{Protein Models} (models that handle protein sequences, e.g., Alphafold~\citep{jumper2021highly}, ESM 2~\citep{lin2023evolutionary}, 
AMPLIFY~\citep{fournier2024protein}). In this paper, we focus on Genomic foundation models due to their general strong capability and wide application domains. As illustrated in \Cref{wrap-tab:bfm_tasks}, bio-foundation models are used not only for sequence generation but also for regression and classification tasks. These include applications such as mutational effect prediction, clinical variant interpretation, and related functional genomics analyses.

\paragraph{Data filtering as a safety approach.} Data filtering has been widely adopted during the pretraining stage of open-weight foundation models to mitigate legal and safety risks. In natural language models, it has proven effective in reducing harmful outputs~\citep{anil2023palm}, limiting private information leakage~\citep{korbak2023pretraining}, and mitigating copyright issues~\citep {min2023silo}. Recent studies have further advanced filtering methods by applying harmfulness classifiers~\citep{chen2025enhancing}. Compared to post-training safeguards such as circuit breakers~\citep{zou2024improving}, data filtering has emerged as a more robust approach in language models~\citep{obrien2025deepignorancefilteringpretraining}. In the domain of bio-foundation models, data filtering is likewise employed to reduce misuse risks. For example, Evo 2 excluded eukaryotic viral data during pretraining~\citep{nguyen2024sequence, brixi2025genome} to mitigate potential biosecurity concerns.


\paragraph{Assessing biological dual-use risk for Foundation Models.} Various efforts have been made to explore dual-use risks in bio-foundation models.
For example, Evo 2 assesses such risks during the deployment stage using metrics like perplexity distribution, mutational effect prediction, protein generation success rates, and ancestry bias in eukaryotic viral sequences~\citep{brixi2025genome}; Genebreaker~\citep{zhang2025genebreaker} investigates the ability to induce bio-foundation models to generate eukaryotic viral sequences via inference-time guided search. Beyond bio-foundation models, several studies have examined dual-use concerns in Language Foundation Models. For instance, system cards from OpenAI model families~\citep{openai2025gpt5, openai2025o3} assess the biological dual-use risk through evaluations on long-form bio-risk questions, multimodal virology troubleshooting, ProtocolQA~\citep{laurent2024lab}, and tacit knowledge probing. In terms of malicious fine-tuning risks, \citet{wallace2025estimating} proposes a worst-case estimation approach that stress-tests maximum knowledge gains after fine-tuning with adequate compute budgets. However, to the best of our knowledge, there is a lack of research assessing the dual-use risks associated with fine-tuning bio-foundation models -- a gap this paper aims to address.



%% file: tables/bfm_tasks.tex
\begin{wraptable}{r}{0.5\textwidth}
\vspace{-18mm} 
\caption{Bio-foundation models can be used for sequence generation and regression/classification tasks. Adversaries can attack the model during both the deployment and fine-tuning stages. However, the risks associated with most attack–task combinations remain underexplored.}
\label{wrap-tab:bfm_tasks}
\resizebox{\linewidth}{!}{
\begin{tabular}{ccc}
\toprule  
 & \textbf{Sequence}& \textbf{Regression /} \\
 & \textbf{Generation} & \textbf{Classification} \\
\midrule
\textbf{Deployment} & \multirow{2}{*}{\cmark~\citep{zhang2025genebreaker}} & \multirow{2}{*}{\textcolor{red}{?} (\Cref{subsec:mutation_pred}, \ref{subsec:vir_pred})}\\  
\textbf{Stage} & & \\
\midrule
\textbf{Fine-Tuning} &\multirow{2}{*}{\textcolor{red}{?} (\Cref{subsec:inter-species-gen})}&\multirow{2}{*}{\textcolor{red}{?} (\Cref{subsec:mutation_pred}, \ref{subsec:vir_pred})}\\  
\textbf{Stage} & & \\
\bottomrule
\end{tabular}}
\vspace{-5mm}
\end{wraptable} 

%% file: sections/benchmark_intro.tex
\section{Evaluating Bio-Foundation Models' Harmful Capabilities}
\label{sec:benchmark_intro}


\input{tables/benchmark_detail}

\subsection{Threat Model}
\label{subsec:threat_model}
 Following the insights from the prior work~\citep{brixi2025genome, yin2023vipal}, we consider a threat model where adversaries seek to exploit open-weight bio-foundation models to design pathogenic eukaryotic viruses. From the adversaries' perspective, the objective is to leverage the model to facilitate pathogen engineering. Specifically, adversaries may (1) generate novel and viable viral sequences; (2) perform targeted \textit{in silico} optimization to enhance hazardous traits such as protein stability or receptor binding affinity, and (3) rank and select potential candidates based on the predicted virulence. Full access to model weights further enables adversaries to fine-tune the model on public biological datasets or probe hidden representations to improve performance on these tasks. From the defender's perspective, the objective is to minimize the risk of such misuse after releasing the model weights. To this end,  pre-release safety approaches will be applied to constrain malicious utility while maintaining scientific value for legitimate users. Our safety evaluations are thus designed to assess the model’s harmful capabilities across the adversarial tasks outlined above.

\subsection{\eval}

Under the threat model in \Cref{subsec:threat_model}, we introduce \eval, a framework that evaluates bio-foundation models' harmful capabilities along three dimensions: Sequence modeling capability (\evalgen), mutational effect prediction (\evalmut), and virulence prediction (\evalvir). We provide an overview of our evaluation framework in \Cref{tab:benchmark_info} and discuss each perspective below (See \Cref{app:dataset_details} for more details).

\paragraph{Sequence Modeling.}To assess the general sequence modeling capabilities on human-infective eukaryotic viruses, we compute the model's perplexity on these genomic sequences, as perplexity quantifies how confidently a model generates sequences within this domain. We collect data from the National Center for Biotechnology Information (NCBI) Virus Repository\footnote{\url{https://www.ncbi.nlm.nih.gov/labs/virus/vssi}} and only keep the sequences with the human host tag (\textit{``Homo sapiens''}). After exact-match deduplication, we obtain 10,247,388 unique sequences, covering 57 families, 131 Genera, and 1,795 Species. This supports a fine-grained analysis of perplexity distribution across taxonomic levels, from individual species to viral families.


\paragraph{Mutational Effect Prediction.} We evaluate mutational effect prediction using 16 Deep Mutational Scanning (DMS) datasets~\citep{fowler2014deep} on human-infecting viruses from ProteinGym~\citep{notin2023proteingym}. Each DMS dataset consists of mutant variants with experimentally measured fitness scores obtained through high-throughput selection assays (e.g., growth, binding, or expression). Fitness scores are estimated by comparing pre- and post-selection sequencing counts, normalized to the wild type. We assess models by computing the absolute Spearman rank correlation $|\rho|$ between the experimental fitness values and the model-derived sequence scores for the same mutants. A larger $|\rho|$ indicates better agreement with experimental rankings and thus a stronger ability to capture mutational effects. We convert ProteinGym's protein sequences into nucleotide sequences for evaluations with Evo 2, in a process detailed in 
\cref{app:nucleotide}. More details on the DMS datasets can be found in \cref{tab:dms_dataset_detail}.



\paragraph{Virulence Prediction.} To evaluate the model's capability on predicting virulence, we use the median lethal dose (\ld) as our metric, which represents the dose required to kill half of a tested population after a specified test duration~\citep{casadevall2017pathogenic}. For consistency, we focus on \ld measurement of Influenza A viruses in mice~\citep{ivan2019rule, yin2023vipal}, and only keep the record from the BALB/C host strain. Since each Influenza A virus strain consists of eight RNA segments, we concatenate them into a single sequence during evaluation.  In our experiment, we evaluate the model's predictive performance by probing hidden-layer representations and measuring the correlations between predicted and observed values using Pearson Correlation.

%% file: tables/benchmark_detail.tex

\begin{table}[t]
\centering
\caption{\eval assess bio-foundation models' harmful capabilities on eukaryotic viral sequences from three dimensions, offering a comprehensive evaluation on the misuse risk.}
\resizebox{\linewidth}{!}{
\begin{tabular}{ccccc}
\toprule
Dataset Name & Eval Capability                  & Source                     & \# Examples  & Metric                         \\
\midrule
\eval-\textsc{Gen} & Sequence Modeling         & NCBI Virus Repository            &    10,247,388                   & Perplexity                     \\
\eval-\textsc{Mut} & Mutational Effect Prediction & 16 Human Virus DMS Datasets                &    156,178              & $|$Spearman $\rho|$ \\
\eval-\textsc{Vir} & Virulence Prediction        & Influenza A Virulence Info &    369                 & Pearson                    \\
\bottomrule
\end{tabular}}
\label{tab:benchmark_info}
\end{table}

%% file: sections/exp_results.tex
\section{Experimental Results}
\label{sec:exp_results}

Under \eval framework, we seek to answer the research questions in \Cref{sec:intro}. Specifically, we first employ \evalgen to assess the extent to which fine-tuning on eukaryotic viral sequences can generalize sequence modeling capability after data filtering (\cref{subsec:inter-species-gen}). We then examine the model's ability to predict mutation effectiveness on \evalmut, comparing pre- and post-fine-tuning performance through both output log-probability and probes on hidden-layer representations (\Cref{subsec:mutation_pred}). Finally, we evaluate the model's capacity to predict virulence by probing the hidden-layer representations before and after fine-tuning. (\Cref{subsec:vir_pred}).

We use Evo2-7B model~\citep{brixi2025genome}, a genome foundation model designed for both nucleotide sequence modeling and biological representation prediction, as our primary model due to its strong performance across diverse downstream tasks. Moreover, its deliberate exclusion of eukaryotic viral sequences during pre-training enables us to assess the effectiveness of data filtering, offering a controlled safety baseline for our experiments. See \Cref{app:exp_details} for more details.

\subsection{Fine-tuning Exhibits Inter-Species Generalizability in Sequence Modeling}

\label{subsec:inter-species-gen}


We test fine-tuning generalization in sequence at two taxonomic scales that differ in evolutionary distance. In the \textit{inter-species} setting (\Cref{subsubsec:inter-species}), one species within a genus is held out to test whether fine-tuning on the remaining species transfers effectively. In the \textit{inter-genus} setting (\Cref{subsubsec:inter-genus}), one genus within a family is held out to examine transfer across a larger evolutionary gap. These two tests indicate how far fine-tuning can bridge evolutionary distance and whether filtering by species or genus meaningfully limits capability recovery. From an adversarial perspective, this also shows how efficiently fine-tuning can be applied, since only sequences beyond the generalizable evolutionary distance need to be preserved in the training set.

\input{tables/inter_species_setup}

\begin{figure}[t]
    \centering
    \begin{subfigure}{\textwidth}
        \centering
        \includegraphics[width=\linewidth]{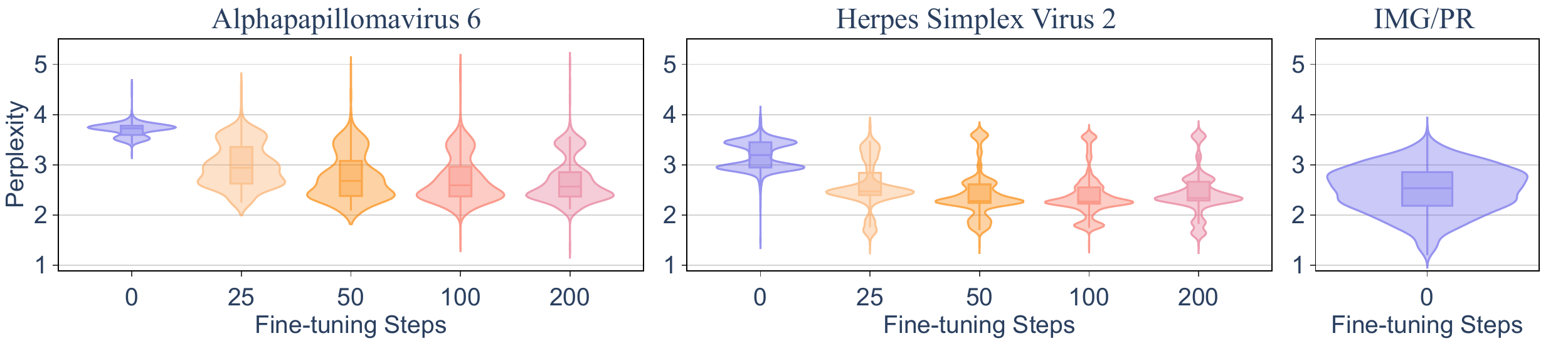}
        \caption{Inter-Species Generalization: Fine-tuning Evo2-7B on all species in Alphapapillomavirus excluding type 6 (left), and all species in Simplexvirus excluding HSV-2 (middle). Right: Baseline perplexity distribution on benign plasmid sequences from IMG/PR.}
    \label{fig:ppl_violin_interspecies}
    \end{subfigure}
    \begin{subfigure}{\textwidth}
    \centering
    \includegraphics[width=\linewidth]{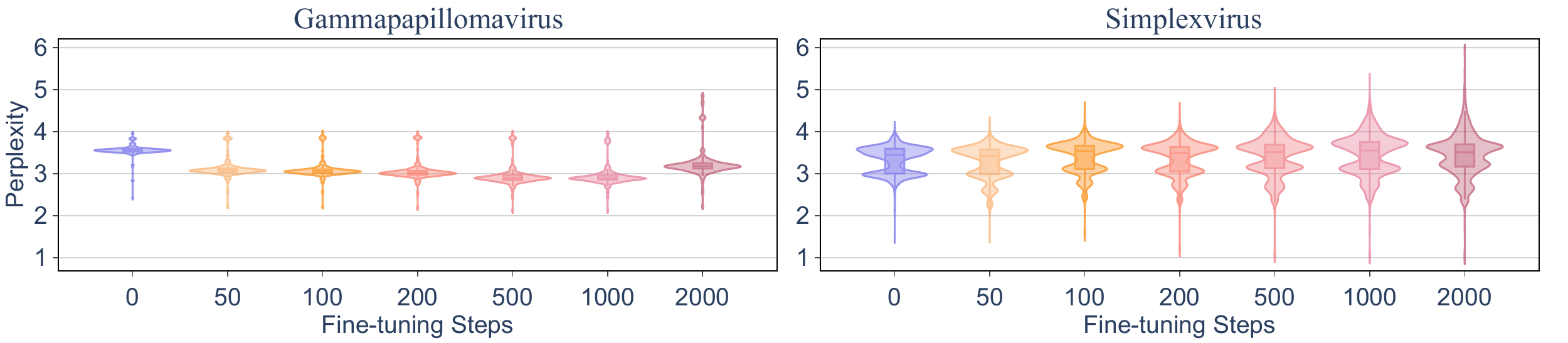}
    \caption{Inter-Genus Generalization: Fine-tuning Evo2-7B on all genera in Papillomaviridae excluding Gammapapillomavirus (left), and all genera in Orthoherpesviridae excluding Simplexvirus (right).}
    \label{fig:ppl_violin_intergenus}
    \end{subfigure}
    \caption{We test whether fine-tuning can show (a) inter-species generalizability, and (b) inter-genus generalizability. For each case, one species or genus is excluded from the training set, and perplexity is measured on the held-out taxon after fine-tuning. Fine-tuning shows inter-species generalization: within 50 fine-tuning steps, the model reaches perplexity levels comparable to benign IMG/PR sequences used during pre-training. In contrast, inter-genus generalization is harder to achieve.}
    \label{fig:ppl_violin}
    
\end{figure}

\subsubsection{Inter-Species Generalizability}
\label{subsubsec:inter-species}

\paragraph{Setup.} We evaluate the inter-species generalizability of fine-tuning on eukaryotic viral sequences through two case studies, summarized in the first two rows of \Cref{tab:inter_species_setup}. The first focuses on Alphapapillomavirus, a genus that will cause cervical cancer and genital warts; The second examines Simplexvirus, which causes skin vesicles and mucosal ulcers. For both cases, we fine-tune the model for 25 to 200 steps, and exclude one species from the training set. Generalization is assessed by computing the perplexity distribution on the hold-out species. All the training and hold-out examples are selected from \evalgen. As a baseline, we also compute the perplexity distribution on a subset of IMG/PR\footnote{\url{https://huggingface.co/datasets/arcinstitute/opengenome2/blob/main/fasta/plasmids_phage/imgpr.fasta.gz}}, the dataset used to pre-train the Evo2-7B model, which consists of benign plasmid sequences.

\paragraph{Observations.} We plot the perplexity distribution across various fine-tuning steps in \Cref{fig:ppl_violin_interspecies}. Notably, for both case studies, the perplexity distribution on the hold-out species quickly dropped to the same level as on IMG/PR within 50 fine-tuning steps, which is merely 0.72 H100 GPU hours under our experiment settings. This indicates that the model can easily generalize to the other species within the same genus after a few steps of fine-tuning. Given the typically high structural similarity across species in the same genus~\citep{baines2007chapter, van2010evolution, iarc2007human}, such a result is not surprising. However, it highlights that data filtering is not tamper-resistant in this setting: excluding one species does not robustly prevent efficient recovery of capabilities through fine-tuning on related species. Moreover, this also implies that malicious fine-tuning could be significantly streamlined, requiring only a subset of representative species rather than exhaustive coverage.

\subsubsection{Inter-Genus Generalizability}
\label{subsubsec:inter-genus}

\paragraph{Setup.}Building on the settings in \Cref{subsubsec:inter-species}, we extend the dataset from the species level to the genus level, the next higher taxonomic rank. This increases the evolutionary distance between the training and test sets. As detailed in the last two rows of \Cref{tab:fine-tuning-config}, we fine-tune the Evo2-7B on all genera in Papillomaviridae excluding Gammmapapillomavirus, and on all genera in Orthoherpesviridae excluding Simplexvirus. After fine-tuning for 50 to 2,000 steps, we evaluate perplexity on the hold-out genera.

\paragraph{Observations.} \Cref{fig:ppl_violin_intergenus} shows that even within the same family, fine-tuning yields lower generalizability across genera compared to across species. In the Papillomaviridae experiments, while fine-tuning for 1,000 steps reduces average test-set perplexity by approximately 20\%, the values remain higher than those on benign sequences from IMG/PR. Similarly, for Orthoherpesviridae, fine-tuning over 2,000 steps produced no significant decrease in perplexity on the held-out genus. Overall, as we move from the species level to the genus level, achieving generalization through fine-tuning becomes more difficult and computationally demanding.



\subsection{Fine-Tuning and Probing Help Recover Mutational Effect Knowledge}
\label{subsec:mutation_pred}

\subsubsection{Log-Likelihood-Based Mutational Effect Prediction}
\label{subsubsec:zero-shot-dms}

\textbf{Setup.} Following prior work~\citep{meier2021language, notin2023proteingym}, we score mutational effects with Evo2-7B model by calculating log-likelihood $\mathcal{S}_{LL} = \sum_{j=1}^{L} \log P(x_j^{mt} | x_{<j}^{mt}, \theta)$ for each mutation $x^{mt}$. Model predictions are compared against experimental DMS scores using Spearman correlation $\rho$, where higher absolute correlation indicates stronger agreement with measured mutational predictive power. To investigate the improvement through fine-tuning, we fine-tune the model on the longest sequences for each species in \evalgen for 100 to 2,000 steps (See \Cref{app:dataset_details} for more details). As baseline, we also include zero-shot scoring with ESM 2, a BERT-style protein language model trained \textit{without} filtering eukaryotic viral sequences~\citep{lin2023evolutionary}. For ESM 2, we adopt masked marginal scoring, which is defined as $\mathcal{S}_{MM}=\sum_{i\in M}\log p(x_i=x_i^{mt}|x_{-i}^{mt})- \log p(x_i=x_i^{wt}|x_{-i}^{wt})$, where $M$ is the set of mutation positions, $x_{-i}$ is the sequence with place $i$ masked, $x^{mt}$ is the mutation sequence, and $x^{wt}$ is the wildtype sequence.



\begin{figure}[t]
    \centering
    \begin{subfigure}{\textwidth}
        \centering
        \includegraphics[width=\linewidth]{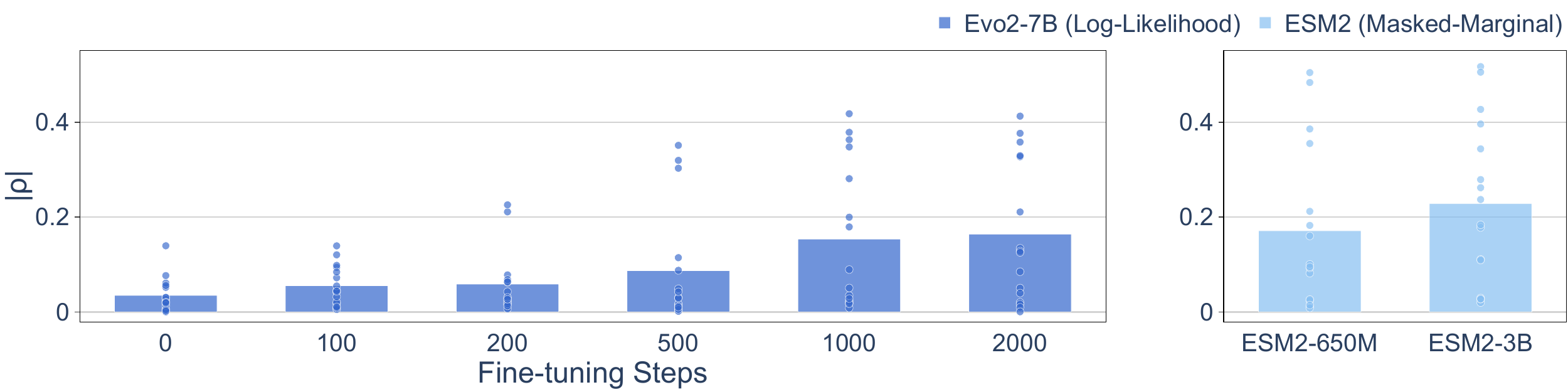}
        \caption{Mutational Effect Prediction on \evalmut.}
    \label{fig:dms}
    \end{subfigure}
    \begin{subfigure}{\textwidth}
    \centering
    \includegraphics[width=\linewidth]{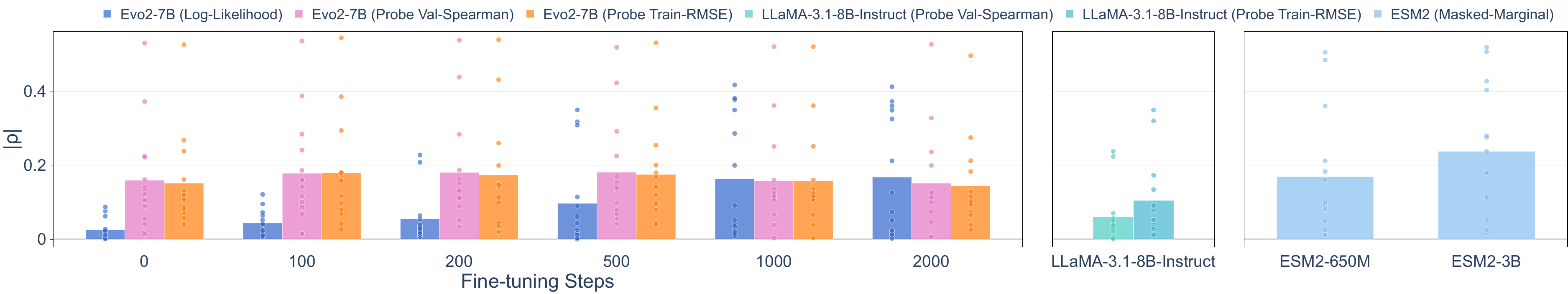}
    \caption{Mutational Effect Prediction on \evalmutprobe}
    \label{fig:dms_probe}
    \end{subfigure}
    \caption{Within 2,000 steps (28.9 H100 GPU Hours), fine-tuning Evo2-7B can achieve a comparable mutational effect prediction as ESM 2 model on (a) \evalmut and (b) \evalmut-\textsc{Probe}.  On \evalmutprobe, even without further fine-tuning, probing the hidden layer representations with the lowest train root mean square error or highest validation $|\rho|$ from Evo2-7B can also achieve a comparable performance as the model without data filtering.}
    \label{fig:dms_full}
    
\end{figure}


\textbf{Observations.} Through \Cref{fig:dms}, we observe that log-likelihood-based mutational effect prediction improves steadily as we fine-tune Evo2-7B, with later checkpoints approaching a similar $|\rho|$ achieved by ESM 2. Previously, Evo 2's low mean $|\rho|$ was reported as evidence of its limited knowledge of mutational effects in human viruses. Here, by fine-tuning the model for 2,000 steps, which is 28.9 H100 GPU hours in our experiment setting, we are able to raise the mean $|\rho|$ from 0.034 to 0.164, which greatly narrows the gap between Evo 2 and other bio-foundation models trained without data filtering.




\subsubsection{Predicting Mutational Effect with Linear Probe}
\label{subsubsec:dms-probe}

\textbf{Setup.} 
To test whether hidden representations encode knowledge relevant for predicting mutational effects, we perform linear probing on model hidden states to predict continuous DMS scores. From \evalmut, we apply stratified sampling based on the DMS score, and sample 500 mutations from each of the 14 DMS datasets (excluding two with fewer than 500 mutations), allocating 400 mutations per dataset for training and 100 for validation, while all remaining mutations are used as the test set. This yields 5,600 training, 1,400 validation, and 148,505 test mutations (denoted as \evalmutprobe, see \Cref{tab:dms_train_test_split} for more details), with probes fit on only 3.6\% of the DMS data and evaluated on 95.4\%.  Here, the probes are \textit{universal} and are trained across all pooled datasets using the closed-form solution with a mean square error objective. Similar to \Cref{subsubsec:zero-shot-dms},  the performance is measured by the absolute Spearman correlation $|\rho|$. Since fine-tuning will alter the hidden representations, we do not predefine a probe layer; instead, for each checkpoint, we train probes on all layers and select (i) the layer with the lowest training root mean square error (RMSE) and (ii) the highest validation Spearman correlation, and report the corresponding test performance in terms of $|\rho|$. For comparison, we not only compute log-likelihood-based mean $|\rho|$ on the test set of \evalmutprobe for Evo2-7B and ESM 2 checkpoints, but also probe the hidden-layer feature from LLaMA-3.1-8B-Instruct~\citep{dubey2024llama}, a natural language model that is not trained on the nucleotide sequences, and shares the same number of layers and hidden layer dimension with Evo2-7B. By comparing the performance between Evo2-7B and LLaMA-3.1-8B-Instruct, we will know how much performance uplift adversaries can achieve with the assistance of bio-foundation models.

\textbf{Observations.}  We present our results on \evalmut-\textsc{Probe} in \Cref{fig:dms_probe}.
Interestingly, even without further fine-tuning, linear probing on Evo2-7B achieves a Spearman correlation of 0.159 when selecting by best validation Spearman, and 0.151 when selecting by train RMSE, which is comparable to the performance of ESM2-650M.  Compared with the best probing result on LLaMA-3.1-8B-Instruct, Evo2-7B still shows a 52.8\% improvement.
Moreover, across our fine-tuning search space, additional fine-tuning does not substantially enhance the expressiveness of hidden-layer representations, as indicated by the nearly unchanged $|\rho|$ values with additional training steps. In contrast, when mutational effects are predicted using log-likelihood scores, the initial $|\rho|$ remains low and follows the same growing trend observed in \Cref{subsubsec:zero-shot-dms} across fine-tuning checkpoints. These findings, together with the small training set size, suggest that the relatively high $|\rho|$ achieved through linear probing is not due to distributional differences between \evalmut and \evalmut-\textsc{Probe}, but rather reflects knowledge already encoded in the model’s hidden representations.



    


\subsection{Latent Virulence Knowledge Persists Despite Data Filtering}
\label{subsec:vir_pred}
\begin{figure}[t]
    \centering
    \includegraphics[width=\linewidth]{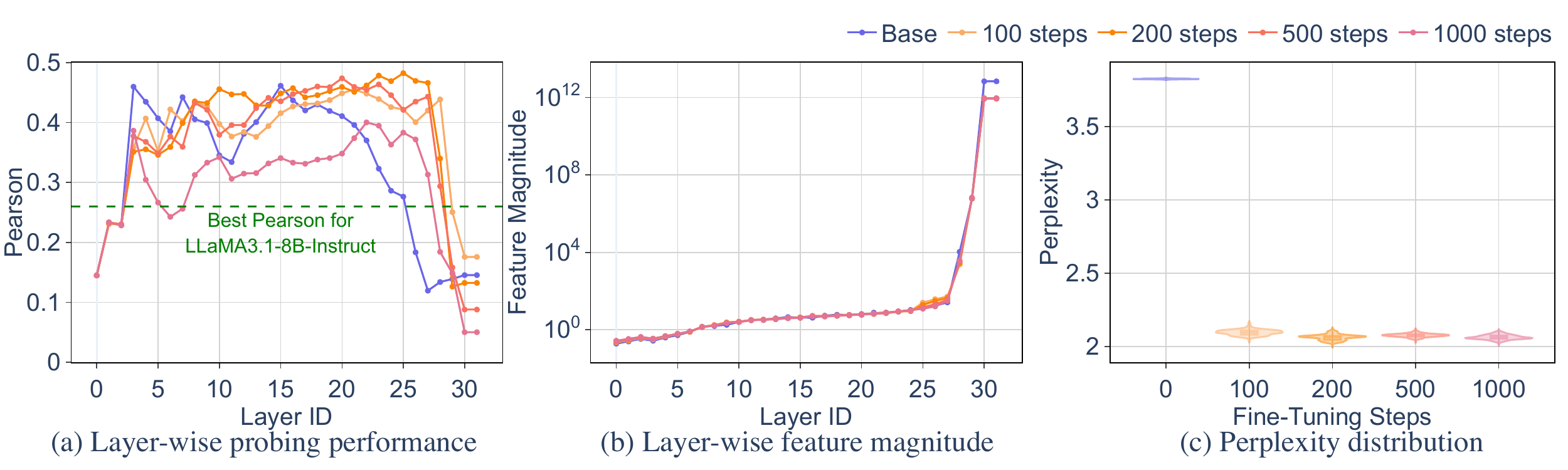}
    \caption{(a) Layer-wise probing results on virulence prediction using \evalvir. Compared with the best probing result from LLaMA-3.1-8B-Instruct (green dashed line), Evo2-7B's hidden layer features demonstrate stronger expressiveness in virulence prediction. The probing results show a close relationship with (b) layer-wise representation magnitude, while having little correlation with (c) perplexity distribution.}
    \label{fig:virulence_prediction}
\end{figure}

\paragraph{Setup.} We assess the model's capability for virulence prediction by probing its hidden layer representation using a linear probe. To train the probe, we stratify and sample 10\% of examples (37 Examples) from \evalvir as the training set, using the remaining 90\% (332 Examples) as the test set. During training, we feed the training examples into the checkpoints and extract hidden-layer representations from the end of a specific Hyena Block. Since the \ld label is continuous and the probe does not include any non-linear functions, we directly compute the closed-form solution, and then evaluate the probe's performance using the corresponding hidden-layer representations from the test set.

Since \evalvir only contains the virulence data for Influenza A virus, we further fine-tune the Evo2-7B model on the influenza A virus sequences from the NCBI Virus Repository. To balance efficiency and diversity, we select the longest sequence per strain and apply stratified sampling based on sequence length, resulting in a training set of 93,844 examples. We fine-tune the model for 100 to 1,000 steps and evaluate layer-wise probing performance across checkpoints.

\paragraph{Observations.} \Cref{fig:virulence_prediction}a illustrates the layer-wise probing analysis reveals that even the base Evo2-7B model achieves relatively high Pearson correlation coefficients across several layers, with a maximum of 0.46, indicating that its hidden representations may already encode strongly predictive values.
To ensure these results are not due to chance, we conduct an ablation study by running the same probing experiment on the LLaMA-3.1-8B-Instruct model, and plot its highest Pearson coefficient across all layers in a green dashed line in \Cref{fig:virulence_prediction}(a). The 77\% performance improvement of Evo2-7B over LLaMA-3.1-8B-Instruct suggests implies that the model has likely acquired harmful knowledge in the latent space during pre-training, despite data filtering. While \citet{Lu2025.09.05.674459} suggested that genomic heterogeneity can inflate the model's performance in binary prediction of human single-nucleotide variants, we argue that this does not directly apply here since we focus on predicting the continuous \ld over entire viral genomes. In fact, the ability to generalize across underlying semantic features in a held-out set is precisely what raises dual-use concerns, as we demonstrate that such emergent knowledge can be introduced at a minimal cost without any further fine-tuning.

Meanwhile, fine-tuned checkpoints offer only marginal gains over the base model -- yielding 5\% increase in maximum Pearson correlation after 200 fine-tuning steps. Moreover, \Cref{fig:virulence_prediction}d shows that the perplexity drops significantly within 100 steps of fine-tuning, yet this does not appear to strongly influence virulence prediction performance, indicating that perplexity may not be a suitable proxy for latent virulence encoding.

To understand why the representation expressiveness diminishes after layer 28, we analyze the layer-wise representation magnitude in \Cref{fig:virulence_prediction}b. We observe the drastic growth of representation magnitude beyond this layer, which may explain the observed collapse. See \Cref{app:evo2_analysis} for more discussions.

%% file: tables/inter_species_setup.tex
\begin{table}[h]
\centering
\caption{Overview of the dataset used for validating inter-species and inter-genus generalizability.}
\resizebox{\linewidth}{!}{
\begin{tabular}{cccccc}
\toprule
\multicolumn{2}{c}{\textbf{Training set}}                    & \multicolumn{2}{c}{\textbf{Hold-Out Set}}    & \multicolumn{2}{c}{\textbf{Baseline}}                     \\
\midrule
\textbf{Species}                            & \textbf{\# Examples} & \textbf{Species}               & \textbf{\# Examples} & \textbf{Dataset}                 & \textbf{\# Examples}            \\
\midrule
Alphapapillomavirus excluding type 6 & 17,055      & Alphapapillomavirus 6 & 569         & \multirow{5}{*}{IMG/PR} & \multirow{5}{*}{5,000} \\
Simplexvirus excluding HSV-2        & 2,618       & HSV-2                 & 1,287       &                         &                        \\
\cmidrule{1-4}                                      
\textbf{Genus}                            & \textbf{\# Examples} & \textbf{Genus }              & \textbf{\# Examples} &                 &             \\
\cmidrule{1-4}
Papillomaviridae excluding Gammapapillomavirus & 17,761              & Gammapapillomavirus & 220 & &  \\
Orthoherpesviridae excluding Simplexvirus      & 8,111                & Simplexvirus      & 3,905 &&                                                    \\
\bottomrule
\end{tabular}}
\label{tab:inter_species_setup}
\end{table}

%% file: sections/conclusion.tex
\section{Discussion}

\paragraph{Conclusion.}In this study, we introduce \eval framework, which offers a comprehensive assessment of the dual-use risks in bio-foundation models across three dimensions. Our experiments with Evo 2 reveal that data filtering is not tamper-resistant under certain circumstances, and may fail to prevent the model from learning malicious capabilities like mutational effect prediction, virulence prediction in the latent space during the pre-training stage. That said, our intention is not to assert that data filtering is completely ineffective, but to highlight the scenarios where it may fall short—underscoring the risks of relying on it as the sole defense mechanism. These findings call for more robust safety strategies for open-weight bio-foundation models that go beyond data filtering alone. We further argue that future safety to open-weight bio-foundation models should account for adversarial manipulations, such as fine-tuning and probing, which are practical for adversaries and can potentially expand the “bubble of risk”~\citep{wei2025dynamic}.


\paragraph{Limitations.}While our work is an initial effort to systematically assess the dual-use risk for bio-foundation models, there is room for improvement in future studies. First, due to the limited data availability, we only collected virulence information from the Influenza A virus. While our results suggest that the model acquires some predictive capability, its performance across other viral families remains untested. Additionally, we only evaluate the Evo 2 model, one of the few open-weight models that explicitly employed data filtering for safety. Broader generalizability requires testing additional models as they become available. Lastly, while our framework covers three critical safety dimensions, other harmful capabilities, such as viral protein sequence generation or host range, remain unexplored and need further investigation.


  

%% file: sections/ethics_statement.tex
\section*{Ethics Statement}

Although our study reveals some signals that Evo 2 may acquire harmful capabilities via fine-tuning and probing, our findings do not directly indicate that the model currently possesses capabilities that constitute a real-world threat. For example, while Evo 2 demonstrates improved performance in mutational effect prediction after fine-tuning or probing, the overall correlation remains modest. This suggests that, while the risks are non-negligible, the model’s current capabilities are still limited and not readily exploitable for malicious purposes. Moreover, all datasets and codebases used for inference and fine-tuning in our research were publicly available prior to this work, meaning our study does not expand the threat landscape. In the end, although we pointed out the weakness of data filtering for bio-foundation models, we aim to catalyze the development of more robust safeguards for open-weight bio-foundation models. Therefore, we believe that the benefits of releasing our research outweigh the potential misuse.

%% file: appendix/dataset_detail.tex
\section{Dataset Details}
\label{app:dataset_details}




\subsection{Fine-tuning Datasets Curation Process}

\begin{figure}[h]
    \centering
    \includegraphics[width=\linewidth]{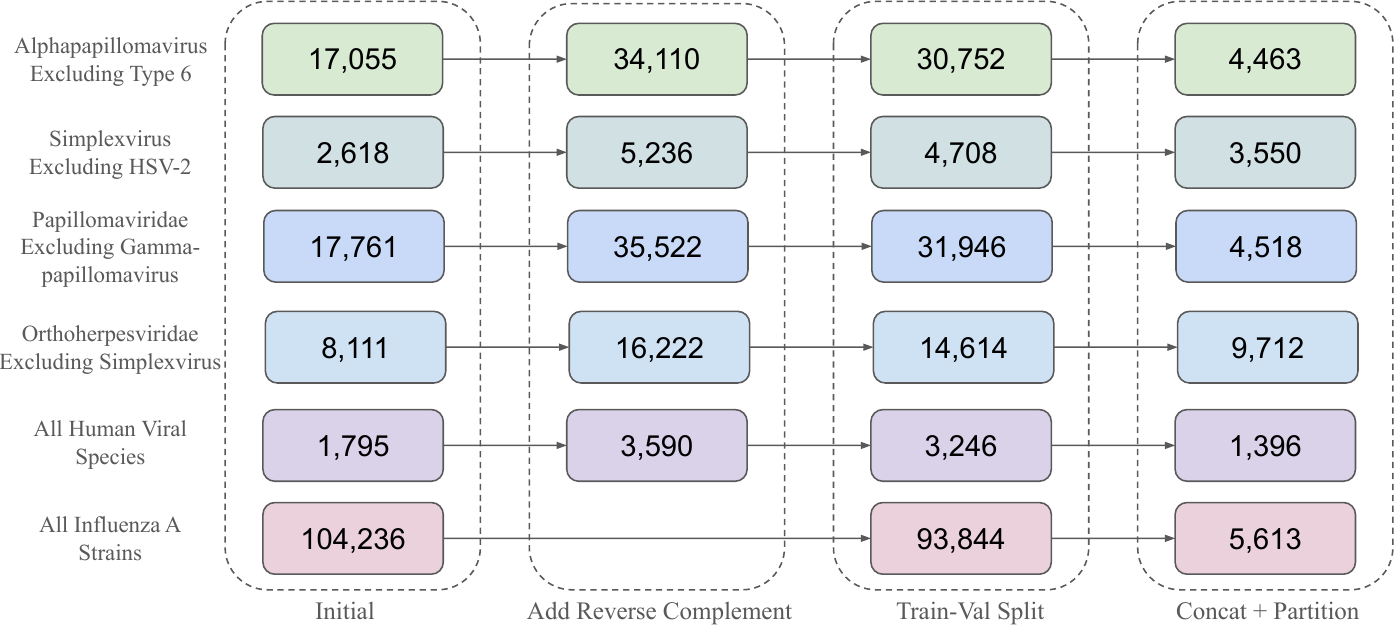}
    \caption{Overview of fine-tuning dataset curation process.}
    \label{fig:ft_dataset_details}
\end{figure}

When curating the fine-tuning dataset, we follow the workflow illustrated in \Cref{fig:ft_dataset_details}, which consists of three stages: \textit{Add Reverse Complement}, \textit{Train-Val Split}, and \textit{Concat+Partition}. For the datasets containing DNA viruses, we first add reverse complement for each sequence to preserve biological symmetry inherent in double-stranded DNA, thereby enhancing generalizability. This process will double the number of examples. For RNA only datasets, like Influenza A dataset, this stage is omitted. Following this, we keep 10\% of examples as the validation set and use the rest 90\% of examples as the training set. In the end, we concatenate all the sequences and then partition them into fixed-length segments of 32,000 tokens. The final number of examples used during fine-tuning is determined after this concatenation and partitioning process. For Papillomaviridae
Excluding Gammapapillomavirus, to avoid potential train-test overlap, we removed all the sequences with empty genus tags. For Orthoherpesviridae Excluding Simplexvirus, to avoid unstable gradient computation, we removed the sequences that contain ``NNN'' (i.e., unknown nucleotides).


\subsection{Protein Sequences to Nucleotide Sequences Conversion Pipeline}
\label{app:nucleotide}
The fitness dataset used are a subset of ProteinGym, which include wild type protein sequence, and various mutations for each DMS experiment. Since Evo 2 is a genomic model, we first convert the protein sequences to nucleotide sequences. Protein to nucleotide sequence conversion is a degenerate process—there are more than one choice of codon for most amino acids. 

We first find wild type nucleotide sequence by using the NBCI BLAST program in three ways: exact match with organism filter, exact match, and 99\% match with seeded random codon replacement for the rest of the amino acids \citep{altschul1990basic}. After the wild type is found or constructed, we perform the mutation swap as detailed by the original DMS protein dataset, and seed the random selection of codon for each replacement at a time.  

Our pipeline do not recreate exact nucleotide sequences that Evo 2 may evaluated on due to the random picking among viable codons, and 2 DMS files that we did not reconstruct due to BLAST not finding wild type nucleotide with high match. We consider this a valid approach because DMS experiments artificially introduce mutations biologically. Furthermore, experimentally we reproduce the mean absolute Spearman rank for human viruses on Evo2-7B, with less than 0.5 mean absolute Spearman rank matching S2B in  \citep{brixi2025genome}.


Additionally, since we are assuming the role of attackers, our approach can be seen as the attacker's best attempt at reconstructing nucleotide sequences and using it to predict the ground truth (and unchanged) DMS score labels.

\subsection{Evaluation Datasets}

\textbf{Zero-shot Mutational Effect Prediction} 
For virus DMS zero-shot fitness reproduction, we use the same DMS listed by the Evo 2 paper in 4.3.6. The original paper mentions 18 datasets and cites 20 studies used for viruses that infects humans. Out of those 20, we use BLAST and seeded random condon selection to create nucleotide counterparts for the DMS protein sequences in ProteinGym. We create nucleotide counterparts for 16 virus DMS datasets. For more details on the conversion, see \ref{app:nucleotide}. The 16 DMS datasets we evaluate on are in \cref{tab:dms_dataset_detail}.

\textbf{Probe-based Mutational Effect Prediction} 
For probing experiments on protein fitness, we sample from the 16 virus protein DMS datasets that have been converted to nucleotides as listed in the section above. We sample 624 balanced samples from each dataset, or the largest balanced subset if there are fewer than 624 samples. We split the samples into balanced 80\% train and 20 \% test. The total training set has a size of 7384, and the total test set has a size of 1868. For ESM2 models, since it takes in protein sequences, we experiment with the original wild type protein sequence, and the protein mutation to calculate masked marginals on the same train and split data. 

\textbf{Virulence Prediction} We use the \ld information for Influenza A viruses from \citet{ivan2019rule} and obtain the corresponding protein sequences from ViPal~\citep{yin2023vipal}\footnote{\url{https://github.com/Rayin-saber/ViPal/tree/main/data}}. Following the conversion pipeline in \Cref{app:nucleotide}, we first filter out entries with missing data and convert the protein sequences into nucleotide sequences. Since each virus consists of 8 genomic segments, we concatenate them into a single sequence. To evaluate the model's ability in predicting virulence, we input the concatenated sequences into the model and extract hidden layer representations for subsequent probing analysis.


\subsection{DMS Datset Overview}

We include details of the 16 human virus DMS datasets in \cref{tab:dms_dataset_detail}. All coarse selection type for the DMS studies are organismal fitness. For each DMS dataset in \evalmut-\textsc{Probe}, we include the corresponding train–val–test split statistics in \cref{tab:dms_dataset_splits}.

\input{tables/dms_dataset_overview}

\input{tables/dms_dataset_probe_overview}

%% file: tables/dms_dataset_overview.tex
\begin{table}[h!]
\centering 
\caption{Summary of Deep Mutational Scanning Datasets used in \evalmut}
\label{tab:dms_dataset_detail}
\resizebox{\linewidth}{!}{
\begin{tabular}{lccc}
\toprule
\textbf{DMS ID} & \textbf{\# Mutants} & \textbf{Sequence Length} & \textbf{Organism} \\ 
\midrule
A0A192B1T2\_9HIV1\_Haddox\_2018 \cite{haddox2018mapping}& 12,577 & 2,556 & HIV-1 (isolate BF520.W14M.C2)\\
A0A2Z5U3Z0\_9INFA\_Doud\_2016 \cite{doud2016accurate} & 10,715 & 1,695 & Influenza A virus (A/WSN/1933(H1N1)) \\
A0A2Z5U3Z0\_9INFA\_Wu\_2014 \cite{wu2014high}& 2,350 & 1,695 & Influenza A virus (A/WSN/1933(H1N1)) \\
A4D664\_9INFA\_Soh\_2019 \cite{soh2019comprehensive}& 14,421 &  2,277 & Influenza A virus (A/green-winged teal/Ohio/175/1986(H2N1)) \\
C6KNH7\_9INFA\_Lee\_2018 \cite{lee2018deep}& 10,754 & 1,698 & Influenza A virus (A/Perth/16/2009(H3N2)) \\
CAPSD\_AAV2S\_Sinai\_2021 \cite{sinai2021generative}& 42,328 & 2,205 & Adeno-associated virus 2 (AAV-2) (isolate Srivastava/1982) \\
ENV\_HV1B9\_DuenasDecamp\_2016 \cite{duenas2016saturation}& 375 & 2,559 & HIV-1 group M subtype B (strain 89.6) \\
I6TAH8\_I68A0\_Doud\_2015 \cite{doud2015site}& 9,462 & 1,494 & Influenza A virus (A/Aichi/2/1968 H3N2) \\
NRAM\_I33A0\_Jiang\_2016 \cite{jiang2016balance}& 298 & 1,359 & Influenza A virus (A/Wilson-Smith/1933 H1N1) \\
PA\_I34A1\_Wu\_2015 \cite{wu2015functional}& 1,820 & 2,148 & Influenza A virus (A/Puerto Rico/8/1934 H1N1) \\
POLG\_DEN26\_Suphatrakul\_2023 \cite{suphatrakul2023functional}& 16,897 & 2,700 & Dengue virus 2 (DENV-2) (strain Thailand/16681/1984) \\
Q2N0S5\_9HIV1\_Haddox\_2018 \cite{haddox2018mapping}& 12,729 & 2,580 & HIV-1 (isolate BG505.W6M.C2.T332N)\\
R1AB\_SARS2\_Flynn\_2022 \cite{flynn2022comprehensive}& 5,725 & 918 & SARS-CoV-2 (isolate MN908947.3, RefSeq NC\_045512.2) \\
RDRP\_I33A0\_Li\_2023 \cite{li2023deep}& 12,003 & 2,271 & Influenza A virus (A/Wilson-Smith/1933 H1N1) \\
REV\_HV1H2\_Fernandes\_2016 \cite{fernandes2016functional}& 2,147 & 348  & HIV-1 group M subtype B (isolate HXB2) \\
TAT\_HV1BR\_Fernandes\_2016 \cite{fernandes2016functional}& 1,577 & 258 & HIV-1 group M subtype B (isolate BRU/LAI) \\
\bottomrule
\end{tabular}}
\end{table}

%% file: tables/dms_dataset_probe_overview.tex
\begin{table}[htbp]
\centering
\caption{Summary of Train--Val--Test Splits in \evalmut-\textsc{Probe}}
\label{tab:dms_dataset_splits}
\begin{tabular}{lcccc}
\toprule
\textbf{Dataset} & \textbf{\# Train} & \textbf{\# Val} & \textbf{\# Test} & \textbf{\# Total} \\
\midrule
A0A192B1T2\_9HIV1\_Haddox\_2018 \citep{haddox2018mapping} & 400 & 100 & 12,077 & 12,577 \\
A0A2Z5U3Z0\_9INFA\_Doud\_2016 \citep{doud2016accurate} & 400 & 100 & 10,215 & 10,715 \\
A0A2Z5U3Z0\_9INFA\_Wu\_2014 \citep{wu2014high} & 400 & 100 & 1,850 & 2,350 \\
A4D664\_9INFA\_Soh\_2019 \citep{soh2019comprehensive} & 400 & 100 & 13,921 & 14,421 \\
C6KNH7\_9INFA\_Lee\_2018 \citep{lee2018deep} & 400 & 100 & 10,254 & 10,754 \\
CAPSD\_AAV2S\_Sinai\_2021 \citep{sinai2021generative} & 400 & 100 & 41,828 & 42,328 \\
I6TAH8\_I68A0\_Doud\_2015 \citep{doud2015site} & 400 & 100 & 8,962 & 9,462 \\
PA\_I34A1\_Wu\_2015  \citep{wu2015functional} & 400 & 100 & 1,320 & 1,820 \\
POLG\_DEN26\_Suphatrakul\_2023 \citep{suphatrakul2023functional} & 400 & 100 & 16,397 & 16,897 \\
Q2N0S5\_9HIV1\_Haddox\_2018 \citep{haddox2018mapping} & 400 & 100 & 12,229 & 12,729 \\
R1AB\_SARS2\_Flynn\_2022 \citep{flynn2022comprehensive} & 400 & 100 & 5,225 & 5,725 \\
RDRP\_I33A0\_Li\_2023 \citep{li2023deep} & 400 & 100 & 11,503 & 12,003 \\
REV\_HV1H2\_Fernandes\_2016 \citep{fernandes2016functional} & 400 & 100 & 1,647 & 2,147 \\
TAT\_HV1BR\_Fernandes\_2016 \citep{fernandes2016functional} & 400 & 100 & 1,077 & 1,577 \\
\midrule
\textbf{Total} & \textbf{5,600} & \textbf{1,400} & \textbf{148,505} & \textbf{155,505} \\
\bottomrule
\end{tabular}
\label{tab:dms_train_test_split}
\end{table}

%% file: appendix/exp_details.tex
\section{Experiment Details}
\label{app:exp_details}

\subsection{Hardware Configuration}
We use Amazon p5.48xlarge\footnote{\url{https://aws.amazon.com/ec2/instance-types/p5/}} as our experiment platform, which consists of NVIDIA H100-80GB GPUs and AMD EPYC 7R13 Processor. All the experiments (fine-tune and inference) are done with 4 NVIDIA H100-80GB GPUs under NVIDIA BioNeMo Framework~\citep{john2024bionemo}.
\subsection{Fine-tuning Configuration}

For all the fine-tuning experiments, we use the fine-tuning configuration in \Cref{tab:fine-tuning-config}.
\begin{table}[h]
\caption{Hyperparameter configurations used in our fine-tuning pipeline}
\centering
\resizebox{\linewidth}{!}{
\begin{tabular}{ccccccc}
\toprule
\textbf{LR}    & \textbf{Optimizer}             & \textbf{LR scheduler} & \textbf{Weight Decay} &\textbf{Warmup Ratio} & \textbf{Batch Size} & \textbf{Seq Length}\\
\midrule
$\text{1.5}\times \text{10}^{\text{-5}}$  & AdamW & Cosine & $\text{1}\times \text{10}^{\text{-3}}$ & 0.05 & 8 &32,000        \\
\bottomrule
\end{tabular}}
\label{tab:fine-tuning-config}
\end{table}

\subsection{Probing Configuration}

In our experiments, we conduct two sets of probing analyses: one on \evalmut using \textit{continuous} \ld labels, and another on bacteriophage sequences using \textit{binary} lifestyle labels.
For the regression task involving continuous \ld labels, we employ a linear probe and compute the closed-form solution directly. For the binary classification task, we train a linear probe with a sigmoid activation function. We use a batch size of 128 and optimize the model using the Adam optimizer.

\subsection{Evaluation Metrics}

\paragraph{Spearman's rank correlation ($\rho$)} When evaluating the model's capability in predicting mutational effect, we use Spearman's rank correlation $\rho$ as our metric. In DMS dataset, each mutant $i$ has a corresponding DMS value $X_i$, based on which we can have a ``groundtruth'' rank $R(X)$. Meanwhile, for each mutant sequence, we can also compute a model-derived (e.g., log probabilities) score $Y_i$, based on which we can have another rank $R(Y)$. Spearman's rank correlation $\rho$ is computed as
\begin{equation}
    \rho = \frac{\text{Cov}\left[R(X), R(Y)\right]}{\sigma_{R(X)}\sigma_{R(Y)}},
\end{equation}
where $\text{Cov}\left[R(X), R(Y)\right]$ is the covariance of rank variables; $\sigma_{R(X)}$ and $\sigma_{R(X)}$ are the standard deviations of the rank variables.

\paragraph{Pearson Coefficient} When probing model's capability in predicting virulence, we use Pearson Coefficient to evaluate the correlation between the groundtruth \ld value $X$ and the predicted value $Y$. The Pearson Correlation is computed as
\begin{equation}
    \text{Pearson} = \frac{\text{Cov}[X, Y]}{\sigma_X\sigma_Y},
\end{equation}
where $\text{Cov}[X, Y]$ is the covariance between $X$ and $Y$; $\sigma_X$ and $\sigma_Y$ are standard deviation.



%% file: appendix/evo2_model_analysis.tex
\section{Evo 2 representation Analysis}
\label{app:evo2_analysis}

In this section, we discuss the architecture and Evo 2 and explain why we observe extremely large representations in the later layers, but can still get reasonable output. Evo 2 uses StripedHyena2~\citep{ku2025systems} architecture, which builds blocks from input-dependent long convolutions. A single Hyena operator creates three streams $q, k, v$ by convolving linear projects of the residual stream $x$ with Toeplitz filters $T, H, K$, then mixes them with inner convolution $G$ and a final projection $M$ to get the output $y$:

\begin{equation}
\label{eq:hyena}
    \begin{split}
    q_t^{\alpha} &= T^{\alpha}_{t t'}\!\left(x_{t'}^{\beta} W^{\beta\alpha}\right),\\
k_t^{\alpha} &= H^{\alpha}_{t t'}\!\left(x_{t'}^{\beta} U^{\beta\alpha}\right),\\
v_t^{\alpha} &= K^{\alpha}_{t t'}\!\left(x_{t'}^{\beta} P^{\beta\alpha}\right),\\
y_t^{\alpha} &= \big(q_t^{\beta}\, G^{\beta}_{t t'}\, k_{t'}^{\beta}\, v_{t'}^{\beta}\big)\, M^{\beta\alpha}.
    \end{split}
\end{equation}

StripedHyena-2 uses three Hyena blocks with different filter parameterizations:
\begin{itemize}
    \item \textbf{Hyena-SE (Short Explicit)}: A short causal depth-wise 1D convolution with explicitly stored tapes per channel. It captures local n-gram-like patterns;
    \item \textbf{Hyena-MR (Medium regularized)}: A medium-length causal filter whose kernel is regularized to keep the frequency response and overall gain controlled.
    \item \textbf{Hyena-LI (Long Implicit)}:  A long-range filter realized implicitly as a small mixture of exponentials evaluated with a stateful recurrence.
\end{itemize}

In Evo2, the layer type is organized following the order of SE-MR-LI, with Rotary Attention Layer inserted in Layer 3, 10, 17, 24, 31. There are a few factors that contribute to the representation explosion in the deeper layers:
\begin{itemize}
    \item \textbf{Missing Layer Normalization on Residual Stream.} There's no explicit layer normalization on the residual stream after adding the block output back. Therefore, if the output of one layer has a slightly larger magnitude than its input and is fed directly into the next, the next layer can then amplify this magnitude further. Over dozens of layers, this may lead to an exponential growth in the values.
    \item \textbf{Input-Dependent and Gated Convolutions.} According to \Cref{eq:hyena}, the filters in Hyena operators are generated from the input sequence itself. Therefore, if the input $x$ already has a large magnitude, the gated filter will also likely have a large magnitude. The final output $y$ is a product of these two large-magnitude tensors, leading to a quadratic increase in magnitude within a single layer. 
\end{itemize}

On the other hand, the RMSNorm in the last layer rescales the representations back before the logits, so the output quality is dominated by the direction instead of the representation magnitude. Therefore, even though the hidden-layer representation explodes in the deeper layers, the model is still able to output meaningful sequences.

To double-check that this phenomenon is not due to the issue from the inference pipeline, we also test the layer-wise representation magnitude distribution using Vortex pipeline\footnote{\url{https://github.com/Zymrael/vortex}}, which is the official inference pipeline used by the Evo2 paper, but still get the same result. This indicates that the representation explosion comes from the model architecture, instead of the inference pipeline.